\documentclass[%
 reprint,
 amsmath,amssymb,
 aps,
]{revtex4-2}
\usepackage[english]{babel}
\usepackage{graphicx}
\usepackage{dcolumn}
\usepackage{bm}

\usepackage[utf8]{inputenc}
\usepackage{graphicx}
\usepackage{subcaption}
\usepackage{hyperref}
\usepackage{amssymb}
\usepackage{amsthm}
\usepackage{amsmath}
\usepackage{xcolor}
\usepackage[version=4]{mhchem}

\begin{document}

\preprint{APS/123-QED}

\title{Mean-field models for morphogenetic processes in physiological contexts}

\author{D. Hernández}
\affiliation{%
	Posgrado en Ciencias de la Complejidad, Universidad Autónoma de la Ciudad de México,%
	San Lorenzo 290, esquina Roberto Gayol, Col. del Valle Sur, Benito Juárez,%
	CDMX, 03100, México
}%

\author{Alejandro Valdés López}
\affiliation{%
	Centro de Investigación y Estudios de Posgrado, Facultad de Ciencias Químicas,%
	Universidad Autónoma de San Luis Potosí, Av. Dr. Manuel Nava 6,%
	Zona Universitaria, 78210, San Luis Potosí, México
}%

\author{E. C. Herrera-Hernández}
\email{erik.herrera@uaslp.mx}
\thanks{Corresponding author.}
\affiliation{%
	Centro de Investigación y Estudios de Posgrado, Facultad de Ciencias Químicas,%
	Universidad Autónoma de San Luis Potosí, Av. Dr. Manuel Nava 6,%
	Zona Universitaria, 78210, San Luis Potosí, México
}%

\keywords{Bio-complexity, reaction-diffusion, heterogeneous media, anomalous diffusion, Turing patterns}

\date{\today}

\begin{abstract}

This work introduces a biophysical formalism to describe the spatiotemporal evolution of the chemical profile in tissues, with the novelty of modeling tissue compartmentalization and the mechanism by which cells maintain the system far from thermodynamic equilibrium via production and/or degradation of substances. The models were derived from conservation laws, chemical kinetic theory and geometric constraints, while considering fundamental properties of tissues to connect theoretical modeling with experimental observations. In a morphogenetic context, each morphogen is described by two coupled reaction–diffusion equations, representing intra- and extracellular dynamics, linked through membrane transport processes such as nonlinear, cross, and anomalous diffusion. We explore the models' morphogenetic potential through diffusion-driven instabilities and discuss how natural tissue heterogeneities influence Turing instabilities and self-organized phenomena. The mathematical structure reveals that two-morphogen systems can produce Turing patterns with multiple characteristic length scales, while the system's dimensionality enables chaotic behavior in well-mixed dynamics. Moreover, due to domain coupling, Turing instabilities are allowed for single-morphogen systems. We used Schnakenberg kinetics to demonstrate that Turing patterns arise even when the activator diffuses faster than the inhibitor ($d<1$), thereby expanding the parameter space for pattern formation. Our results suggest that tissue spatial structure has important consequences for Turing instability mechanisms, in some cases weakening the usual conditions for its emergence while widening the possible patterns it can produce. The proposed framework offers a minimal mathematical basis to explore emergent dynamics in biological and synthetic contexts, with potential applications in developmental biology and tissue engineering.

\end{abstract}

\keywords{Bio-complexity, reaction-diffusion, heterogeneous media, anomalous diffusion, Turing patterns}
\maketitle


\section{Introduction}
\label{sec:intro}

In general terms, biologic and synthetic tissues are cell aggregates that occupy  specific regions of space, and perform different types of spatio-temporal coordinated behaviour at the aggregate level. These modes of collective behaviour {\em emerge} from the interactions of the tissue with the “outside world” (which includes the physiological context of a multicellular organism) and from interactions among its constituents, whether  natural, genetically modified, or synthetic cells \cite{adamala_present_2024}. These interactions are usually a mixture of physico-chemical “signals” which individual cells can interpret and react to, according their genomes and epigenomes \cite{lodish_molecular_2021}. The way cells react normally reduces to the activation/inhibition of the production of specific sets of proteins with some biological function, or that participate in some biological process. For example, enzymes that promotes/inhibits chemical reactions inside a cell, proteins associated with quorum sensing in biofilms \cite{parsek_sociomicrobiology_2005}, or  molecules pertaining to cellular pathways that interprets Bicoid gradients, and Bicoid itself; a set of proteins responsible for establishing positional information for the first developmental steps in the fruit-fly embryo \cite{green_positional_2015}.

Research has shown that large-scale properties of cell aggregates, like resistance to stretching or shearing, or their permeability properties, usually have a common sets of causes with \emph{ self-organized  phenomena} and includes, among other things, the set of molecules  synthesized  by cells, the cells shape and spatial density, their membrane elastic and permeability properties, or the elastic and flow-through characteristics of the surrounding extracellular matrix. These in turn, are a common consequence of two more fundamental facts:

\begin{itemize}
	\item tissues are {\em heterogeneous systems} where {\em intra} and {\em extra-cellular spaces} can be clearly defined, and where each of these regions can be considered as {\em crowded systems} with distinct chemical compositions, different allosteric and excluded volume interactions, and therefore different types of {\em non-linear transport} and {\em anomalous kinetics} \cite{minton_influence_2001,subramanya_molecular_2024,vanag_cross-diffusion_2009};
	\item tissues are  {\em non-equilibrium systems}  maintained this way by the cells agency and energy input; any macroscopic and/or {\em emergent property} associated with them should be modulated or induced by the {\em cells  physico-chemical states}.
\end{itemize}

Based on this, one would naturally infer that  mean-field mathematical models aiming to describe a tissue macroscopic physical or dynamic property, should include the above two points and their consequences. Nevertheless, a large fraction of the literature for this type of research, reduces to the application of homogeneous reaction-diffusion equations (sometimes extended to include mechanic/advective interactions or mild heterogeneities \cite{cruywagen_tissue_1992,baker_partial_2008}), where the heterogeneity and physical structure of the physiological context is obscurely buried in the models parameter set, usually composed of effective diffusion coefficients and kinetic rate constants \cite{murray_mathematical_2003,maini_turings_2012,scholes_comprehensive_2019}. These equations, although successfully applied in different bio-medical and developmental problems (ranging from biofilm dynamics and growth \cite{wang_review_2010}, to self-organized chemical pre-patterns for limb, lung, and root development \cite{murray_mathematical_2003,menshykau_branch_2012,brena--medina_mathematical_2014}), are still crude representations of the actual physiological complexity they try to describe. This jeopardizes our understanding of the potential that living tissues have in terms of collective behaviors, and could miss key features necessary for a broader understanding of their macroscopic/{\em emergent properties}, a basic requirement for their laboratory synthesis and control.

In a developmental context, reaction-diffusion equations are generally classified as Turing or morphogenetic models, and taxis models \cite{baker_partial_2008}. For Turing models, the equations are  used as a framework to study the {\em emergence} of concentration patterns of chemical substances known as morphogens \cite{turing_chemical_1952,gierer_theory_1972,landge_pattern_2020,van_gorder_pattern_2021}. Morphogens are usually synthesized by the tissue cells, and under specific \emph{non-equilibrium conditions} they self-organize as stationary chemical pre-patterns that can be used for positional information in different scenarios, like cell differentiation during development \cite{ferrell_bistability_2012,avila_modelando_2007}. Taxis models on the other hand include the cells density as an extra state variable, and have been uses for the description of spatial structures in bacteria colonies and animal skin pigment patterns \cite{baker_partial_2008,murray_mathematical_2003,wang_review_2010}.

In this work, we develop a formalism from which the spatio-temporal evolution of the chemical profile of different types of tissue can be described. This is carried out using reaction-diffusion models that take into account the fundamental characteristics of tissues as described above. The  models  allow to explore the role that {\em cells physico-chemical states}, along with the tissue {\em heterogeneity} and selective permeability play on different problems {\it e.g.} drug distribution/adsorption in specific physiological settings, biofilm  organization strategies or excitable dynamics, among others \cite{wang_review_2010,avila_modelando_2007}. Here however, we will focus on the models morphogenetic potential within a Turing instability context. 

The reasons for this are threefold. First, during the last decades Turing models have been gradually recognized by the biology community, to represent one of the principal mechanisms (along with positional information {\it a la Wolpert}) of self-organization and morphogenesis in developmental processes \cite{green_positional_2015,wolpert_positional_1989}. Second, as a consequence of this recognition, there is now an active effort within the synthetic biology community to produce and control synthetic Turing patterns \cite{vittadello_turing_2021,davies_using_2017,scholes_three-step_2017}; the type of models herein presented could shed some light into these efforts. Third, their application to pattern formation problems in a physiological context, integrates and complements a series of  findings published in the last couple of decades about the effects of spatial heterogeneity in Turing patterns and instability, the type of pattern their joint effect can produce, and how complex these can be given a number of morphogens. Preliminary analysis of the models supports the claim that Turing patterns are probably a common natural phenomena, and that the necessary conditions for their emergence could actually be weaker than those predicted by classical theory (homogeneous reaction-diffusion equations) \cite{scholes_comprehensive_2019,landge_pattern_2020,krause_modern_2021,rauch_role_2004}.

In Section \ref{sec:mean_field_model}, mean-field models describing the chemical dynamics of different tissues are obtained from general geometric, kinetic, and mass conservation principles. In Section \ref{sec:discussion}, a brief discussion about their mathematical structure and interpretation is given, followed by a comparison with other Turing models that have been used to describe  pigment patterns in different species, and Turing instabilities in structured spatial domains. Furthermore, we show that under specific physiological settings, is possible to have Turing patterns without the usual constrains on the morphogens diffusion coefficients. Finally, some theoretical consequences, future developments and conclusions are presented.

\section{\textbf{Mean-field models}}
\label{sec:mean_field_model}

First, we define what we mean by a {\em cellular field}; a useful concept for the models construction and understanding. \\
{\em Cellular field:} An ordered pair $(\Omega,N)$ such that $\Omega\subset \mathbb{R}^d, d=1,2,3$ and $N\in \mathbb{R}$. Here, $N$ denotes the average number of a given type of average cell, that occupies a spatial domain denoted by $\Omega$.

Consider then a {\em cellular field} where the spatial region occupied by the $i-th$ cell is denoted as $\Omega_i$, and  the {\em extra-cellular space} as $\Omega_o$; it follows that $\Omega=\Omega_o\cup \bigl(\bigcup_i \Omega_i\bigr)$.   Moreover, observe that relative to the aggregate permeability, there are only three possible spatial arrangements for its individual constituents; these will be referred as:

\begin{itemize}
	\item[T1)] {\it Non-percolating tissues:} $\Omega_o$  percolates across $\Omega$ but $\bigcup_i \Omega_i$ does not;
	\item[T2)] {\it Percolating tissues:} $\Omega_o$  does not percolates across $\Omega$, whereas $\bigcup_i \Omega_i$ does;
	\item[T3)] {\it Bi-percolating tissues:} $\Omega_o$ and $\bigcup_i \Omega_i$  percolate across $\Omega$.
\end{itemize} 

\begin{figure}[htbp]
	\centering
	\begin{subfigure}[b]{0.4\textwidth}
		\centering	\includegraphics[trim=8.5cm 3cm 10cm 3cm, clip, width=1\textwidth]{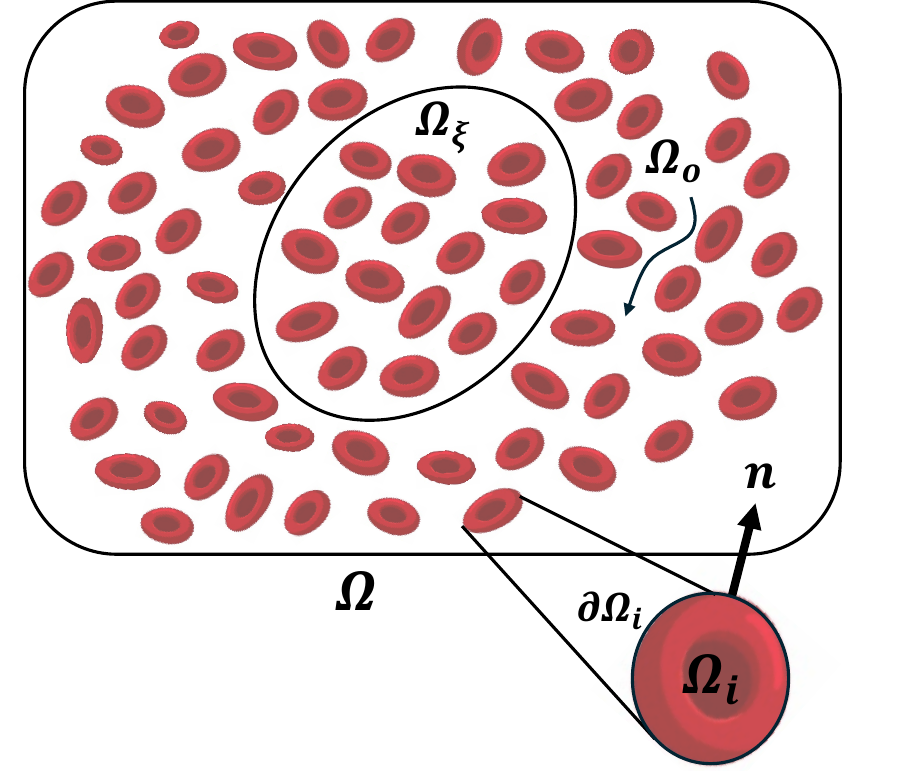}
		\caption{}
		\label{fig:1a}
	\end{subfigure}
	\vskip\baselineskip
	\begin{subfigure}[b]{0.4\textwidth}
		\centering
		\includegraphics[trim=6.5cm 2.5cm 9cm 1cm, clip, width=1\textwidth]{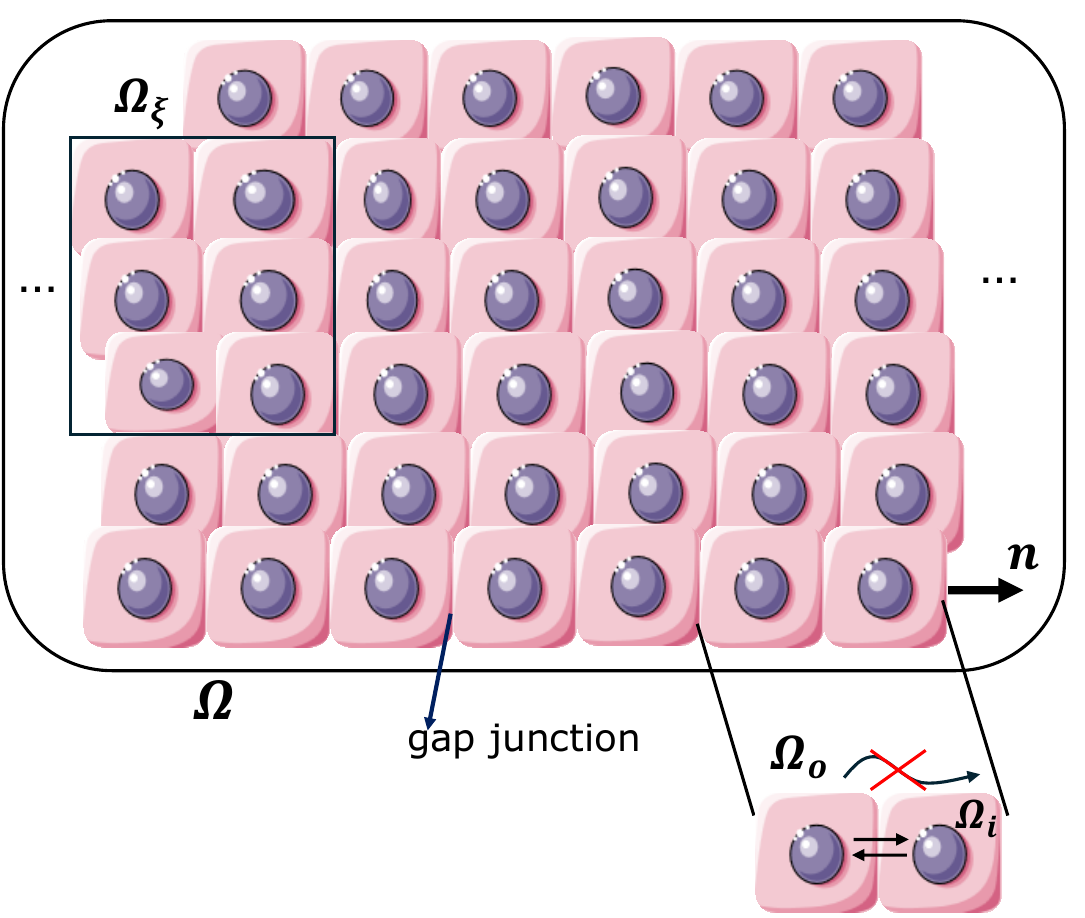}
		\caption{}
		\label{fig:1b}
	\end{subfigure}
	
	\vskip\baselineskip
	
	\begin{subfigure}[b]{0.4\textwidth}
		\centering	\includegraphics[trim=7.5cm 4.5cm 10cm 1cm, clip, width=1\textwidth]{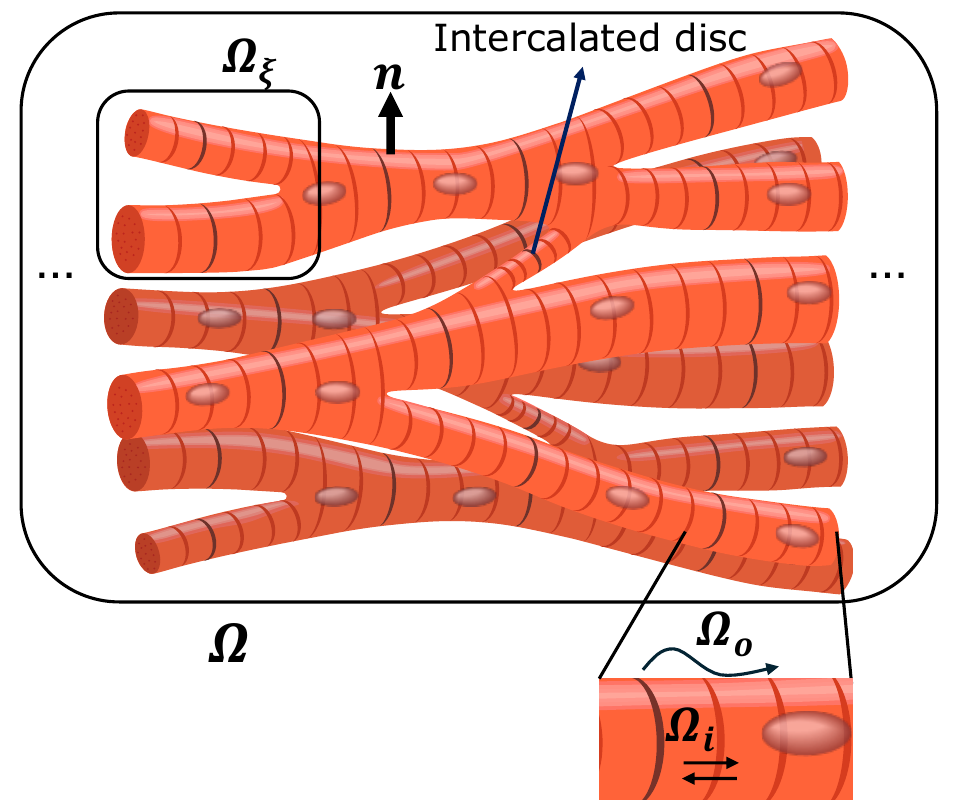}
		\caption{}
		\label{fig:1c}
	\end{subfigure}
	
	\caption{Tissues visual representation. (a) Non-percolating tissue, {\it e.g.} blood; (b) Percolating tissue, {\it e.g.} specific types of epithelial tissue; (c) Bi-percolating tissue, {\it e.g.}  cardiac tissue. In cardiac tissue intercalated discs are specialized structures that includes gap junctions and connect muscle cells \cite{zhao_intercalated_2019}.} 
	\label{fig:1}
\end{figure}

Although not a standard classification, these terms will allow to associate particular model equations with kinds of tissue that share specific  spatial properties, whether natural or man made. Particular cases of \emph{non-percolating} tissues in the Animalia kingdom, are blood and other types of connective tissue where cells are surrounded and spatially separated by an \emph{extra-cellular matrix} and \emph{extra-cellular fluids}. Common cases of epithelial tissue however, can be considered as  \emph{percolating} and \emph{bi-percolating} systems. Here, the interior of adjacent cells is physically/directly connected through microscopic membrane structures called gap junctions \cite{lodish_molecular_2021}. Therefore, long-range transport and cellular communication can be achieved through \emph{intra-cellular domains}, or through \emph{intra} and \emph{extra-cellular domains} (hence the terms \emph{percolating} and \emph{bi-percolating}). See Figure (\ref{fig:1}) for a visual representation of each type of tissue. Notice that from the above discussion, it follows that any tissue has an average cell density given by: $\rho_c=N/Vol(\Omega)$.

We assume that the \emph{physico-chemical state} of any cell inside a tissue, is  determined by the average density of every molecule synthesized and consumed by the cell, in order to perform a biological function or advance some cellular process. If we name those molecules by capital letters: $\{A,B,C,D,...,M\}$, and assume for the moment there are no chemical reactions between them, the temporal evolution of the \emph{chemical state} of any cell in a tissue is given by the following equations:
\begin{equation}\label{eq1}
	\begin{split}
		\frac{d}{dt}\int_{\Omega_i}\rho_{A}d\Omega=\frac{1}{\tau_A}+\sum_{k}\frac{1}{\tau_{A,k}}\int_{\Omega_i}\rho_k d\Omega+f_A(t)\\
		f_A(t)=-\int_{\partial \Omega_i}\Lambda_A \cdot nda\\
		\vdots\\
		\frac{d}{dt}\int_{\Omega_i}\rho_{J}d\Omega=\frac{1}{\tau_J}+\sum_{k}\frac{1}{\tau_{J,k}}\int_{\Omega_i}\rho_k d\Omega+f_J(t)\\
		f_J(t)=-\int_{\partial \Omega_i}\Lambda_J \cdot nda\\
		\vdots\\
		\frac{d}{dt}\int_{\Omega_i}\rho_{M}d\Omega=\frac{1}{\tau_M}+\sum_{k}\frac{1}{\tau_{M,k}}\int_{\Omega_i}\rho_k d\Omega+f_M(t)\\
		f_M(t)=-\int_{\partial \Omega_i}\Lambda_M \cdot nda
	\end{split}
\end{equation} 
where left hand side is the time rate of change for the average number of each molecule, $\rho_J$ denotes the density for $J$-type molecules, and  $1/\tau_J$ their production rate. On the other hand, $1/\tau_{J,k}$  defines the production/degradation rate of $J$ but mediated by a molecule in $\{A,B,C,D,...M\}$. Notice that the cellular processes underlying these terms, are not explicitly considered \footnote{In general, these terms coefficients are density dependent functions.}. Notice also that such processes are responsible for maintaining the tissue \emph{far from thermodynamic equilibrium}. The remaining symbols in the above equation: $\partial \Omega_i$, $f_J(t)$, $\Lambda_J$ and $n$,are respectively a cell external surface, the flux and flux density of $J$-type molecules from or to adjacent \emph{extra-cellular space}, and an unitary normal vector to the cells external surface at a given point.

The \emph{extra-cellular density} for $J$ at a specific spatial location is given  by
\begin{equation}\label{eq2}
	\rho_{J,o}(x,t)=\frac{\int_{\Omega_{\xi,o}}\rho_Jd\Omega}{Vol(\Omega_\xi)}
\end{equation}
where $x \in \Omega_\xi$, and $\Omega_\xi \subset \Omega$ such that: $Vol(\Omega_i)<< Vol(\Omega_\xi)<< Vol(\Omega)$, and where $\Omega_{\xi,o}=\Omega_\xi\cap \Omega_o$.

Consider now \emph{bi} and \emph{non-percolating tissues}, and for simplicity assume that the \emph{extra-cellular space}  is well characterized by effective coefficients of normal diffusion and chemical kinetic processes. Then, the evolution equation for this state variable is given by 
\begin{equation}\label{eq3}
	\frac{\partial \rho_{J,o}(x,t)}{\partial t}=d_{J,o}\nabla^2\rho_{J,o}(x,t)-\gamma_J\rho_{J,o}+g_J(x,t)
\end{equation}
with $d_{J,o}$ the \emph{extra-cellular} effective diffusion coefficient, and $\gamma_J$ an \emph{extra-cellular} degradation rate for this particular molecule and environment. The term $g_J$, denotes the density of $J$-type molecules that leaves or enter the \emph{extra-cellular space} in $\Omega_\xi$ from or to the cells interior and through their membrane, per unit time, and whose mathematical expression is
\begin{equation}
	g_J(x,t)=-\rho_c\int_{\partial\Omega_i}\Lambda_J\cdot n_oda
\end{equation}
where $n_o=-n$, and $n$ is the same unitary vector defined at Equation (\ref{eq1}).

Given the flux density is mediated by the cells membrane, it can be written as a superposition of {\bf ac}tive and {\bf pa}ssive transport processes: $\Lambda_J=\Lambda_{ac}+\Lambda_{pa}$. For active transport, the flux density is usually driven by a free-energy supply from other gradients \cite{hill_free_2004,keener_mathematical_2009}. For passive transport and given we are dealing with \emph{heterogeneous crowded media}, the flux density should have the possibility to include, besides the classical facilitated {\bf n}ormal {\bf d}iffusion \cite{keener_mathematical_2009},  {\bf n}on-{\bf l}inear {\bf d}iffusive, {\bf c}ross-{\bf d}iffusive, and {\bf a}nomalous {\bf d}iffusive phenomena, each of which has been reported in experiments or is expected by theory \cite{minton_influence_2001,subramanya_molecular_2024,vanag_cross-diffusion_2009,bouchaud_anomalous_1990,mendez_reactiontransport_2010}. A proposal on how to accomplish this for $J$-type molecules can be seen in the following expressions:

\begin{equation}\label{pas}
	\begin{aligned}
		\Lambda_{pa}={}&\Lambda_{nd}+\Lambda_{nld}
		+\Lambda_{cd}+\Lambda_{ad},\\
		\Lambda_{nd}={}&-d_{J,m}\nabla\rho_J,\\
		\Lambda_{nld}={}&-d_{J,m}(\rho_J)\nabla\rho_J,\\
		\Lambda_{cd}={}&-\sum_k c_{J,k}(\rho_J,\rho_k)\nabla\rho_k,
		\qquad k\neq J,\\
		\Lambda_{ad}={}&
		-c_{NF,J}
		\Biggl[\int_0^1 \phi(\xi)
		\frac{\partial^{1-\xi}}{\partial t^{1-\xi}}
		\,d\xi\Biggr]\nabla\rho_J\\
		&-c_{MF,J}
		\int_0^t \Xi(t-t')
		\nabla\rho_J\,dt',\\
		\Xi(t)={}&
		\mathcal{L}^{-1}
		\Biggl(
		\Bigl[
		\int_0^1
		\hat{\phi}(\xi)
		s^{\xi-1}
		\,d\xi
		\Bigr]^{-1}
		\Biggr), \\
		\qquad 0&<\xi\le1, \quad
		k\in \{A,B,...,M\}.
	\end{aligned}
\end{equation}

Here, $d_{J,m}$ and $d_{J,m}(\rho_J)$ represent respectively, a facilitated {\bf m}embrane diffusion coefficient and a {\bf m}embrane diffusion coefficient which varies as a function of $\rho_J$. The symbols given by $c_{J,k}(\rho_J,\rho_k)$ and associated with cross-diffusive density fluxes, are real-valued differentiable functions such that: $\lim_{\rho_J\rightarrow 0}c_{J,k}(\rho_J,\rho_k)=0$. The anomalous density flux, a superposition of two subdiffusive processes, and the most general form for including this type of anomalous kinetics, was obtained from the {\bf N}ormal and {\bf M}odified {\bf F}orms of the distributed order fractional diffusion equation. The density flux of the normal form, represented by the term with the symbol $c_{NF,J}$ as a coefficient, describes decelerating subdiffusion; a process where transport becomes less anomalous as a function of time. The modified form on the other hand, represents accelerating subdiffusion, a process where transport becomes more anomalous with time \cite{sandev_distributed-order_2015,sokolov_diffusion_2005}. The coefficients of these two terms are positive constants describing relative weights between subdiffusive processes of $J$ molecules moving through the tissue cells membranes. The functions $\phi$ and $\hat{\phi}$ denote probability densities for the fractional differentiation order, whereas the operators $\partial^{1-\xi}/\partial t^{1-\xi}$ and $L^{-1}(x)$, the Riemann-Liouville fractional derivative of order $1-\xi$ and the inverse Laplace transform. In this expressions,  if $\hat{\phi}(\xi)=\phi(\xi)=\delta(\xi-\xi_0)$, with $\delta(x)$ a Dirac delta function, $\Lambda_{ad}$ reduces to the subdiffusive density flux associated with the usual Riemann-Liouville fractional subdiffusion equation of order $\xi_0$ \cite{metzler_restaurant_2004}.

For the  \emph{intra-cellular media}, the density $\rho_J$ is given by 
\begin{equation}
	\rho_{J,i}(x,t)=\rho_c\int_{\Omega_i}\rho_Jd\Omega
\end{equation}

\begin{figure}[htbp]
	\centering
	\includegraphics[trim=10cm 6.8cm 7.5cm 3.3cm, clip, width=0.5\textwidth]{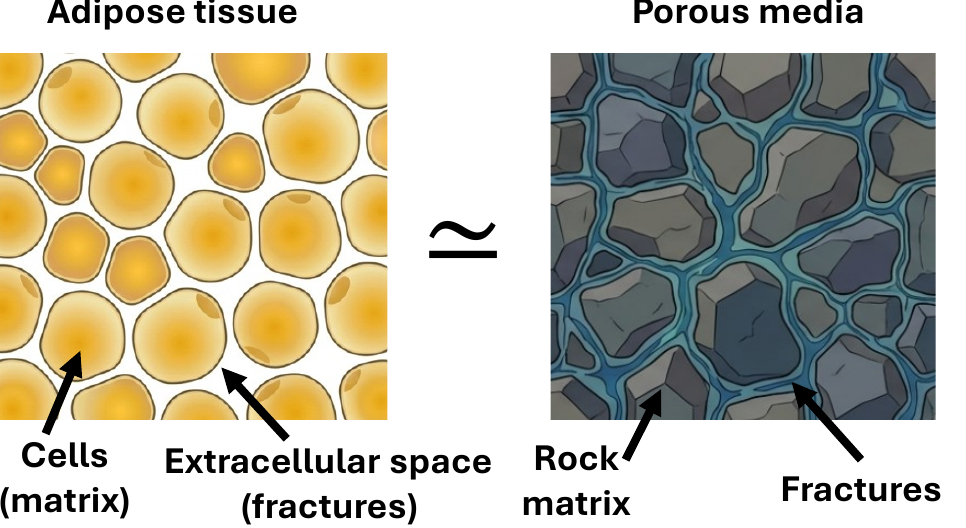}
	\caption{Schematic analogy between a \emph{non-percolating} biological tissue composed of discrete cells and a geological double-porosity medium. Cells play the role of the solid matrix, while the interconnected extracellular space corresponds to the fracture network, which governs transport processes.}
	\label{fig:analogy}
\end{figure}

Here, it is implicitly assumed that $\rho_c = N/Vol{(\Omega)}\approx N'/ Vol{(\Omega_{\xi})}$ with $N'$ the number of cells in  $\Omega_{\xi}$. From the above expression and Eqs.(\ref{eq1}) and (\ref{eq3}), it follows that the evolution equations for $\rho_J$ in \emph{bi-percolating tissues} are:
\begin{equation}\label{mbp}
	\begin{split}
		\frac{\partial \rho_{J,i}}{\partial t}=d_{J,i}\nabla^2\rho_{J,i}+\alpha_J+\sum_{k}\beta_{J,k}\rho_{k,i}
		-\rho_c\int_{\partial \Omega_i}\Lambda_J \cdot nda\\
		\frac{\partial \rho_{J,o}}{\partial t}=d_{J,o}\nabla^2\rho_{J,o}-\gamma_J\rho_{J,o}  +\rho_c\int_{\partial\Omega_i}\Lambda_J\cdot n da
	\end{split}
\end{equation}
where $d_{J,o}$ and $d_{J,i}$, the intracellular diffusion coefficient, are both strictly positive, and where $\alpha_J=\rho_c/\tau_J \in \mathbb{R}^+ $,  $\beta_{J,k}= 1/\tau_{J,k}\in \mathbb{R}$ and $k\in\{A,B,C...M\}$.\\
	\emph{Percolating} and \emph{non-percolating tissues} on the other hand, are described by Eq.(\ref{mbp}) with $d_{J,i}\neq0,d_{J,o}=0$ and $d_{J,i}= 0,d_{J,o}\neq0$ respectively. In Eq.(\ref{pas}),  the flux densities for $J$ molecules are functions whose arguments are morphogen densities and/or morphogen gradients defined at the cells external membrane. Henceforth, the following approximations:

\begin{equation}\label{eqap0}
	\nabla\rho_J|_{x \in \partial \Omega_i}\approx \frac{\rho_{J,o}-\rho_{J,i}}{\Delta l},\quad  \text{and} \quad \rho_J|_{x \in \partial \Omega_i}\approx \frac{\rho_{J,o}+\rho_{J,i}}{2},
\end{equation}

with $\Delta l$ the membrane average width, allow to close the system equations. Interestingly, in  absence of morphogen production and degradation, the model corresponding to \emph{non-percolating tissues} reduces to the double porosity model when only facilitated linear diffusion is considered at the membrane. This model has been widely used in water and oil production \cite{warren_behavior_1963,hernandez_non-local_2021}, and the schematic representation of the porous media in which it is applied, corresponds exactly to what one would expect with these type of tissue, see Figure (\ref{fig:analogy}).   

Until this point, we assumed no chemical reactions. Notice that these are easily included in Eq.(\ref{mbp}) by adding the usual nonlinear expressions obtained from mass action law and chemical kinetic theory \cite{mendez_reactiontransport_2010}. Hence, the temporal evolution for the \emph{physico-chemical state} of any tissue with chemical reactions, is given by:

\begin{equation}
	\label{perc}
	\begin{aligned}
		\frac{\partial \rho_{j,i}}{\partial t}
		={}&d_{j,i}\nabla^2\rho_{j,i}
		+\alpha_j
		+\sum_k \beta_{j,k}\rho_{k,i}
		-\frac{\rho_c s_c}{\Delta l}\hat{\Lambda}_j \\
		&+\eta_{j,i}\!\left(
		\rho_{A,i},\rho_{B,i},\ldots,
		\rho_{j,i},\ldots,\rho_{M,i}
		\right), \\[2mm]
		\frac{\partial \rho_{j,o}}{\partial t}
		={}&d_{j,o}\nabla^2\rho_{j,o}
		-\gamma_j\rho_{j,o}
		+\frac{\rho_c s_c}{\Delta l}\hat{\Lambda}_j \\
		&+\eta_{j,o}\!\left(
		\rho_{A,o},\rho_{B,o},\ldots,
		\rho_{j,o},\ldots,\rho_{M,o}
		\right), \\[2mm]
		&0\le d_{j,i},d_{j,o},
		\qquad
		j,k\in\{A,B,C,\ldots,M\}.
	\end{aligned}
\end{equation}

Here, $s_c$ denotes the average surface area of a cell, and  $\hat{\Lambda}_j$ an approximate flux density obtained from Eq.(\ref{pas}) after substituting the approximations of Eq.(\ref{eqap0}) and factorizing the cells surface area and membrane width. The  functions $\eta_{j,i}$ and $\eta_{j,o}$, denote the chemical kinetics for each physiological domain and particular set of molecules, and comply with the following restrictions: $\sum_j \eta_{j,o}=\sum_j \eta_{j,i}=0$.

Contrary to classical reaction-diffusion theory, in Eq.(\ref{perc}) there are two reaction-diffusion equations for each morphogen. \emph{This is a direct consequence of including spatial heterogeneity in the models, and implies that the dynamic possibilities for any set of morphogens are amplified in a tissue}. For example, the chemical kinetics of a two morphogen system in any tissue, is described by four non-linear ordinary differential equations, opening the possibility for otherwise prohibited chaotic behaviors. Moreover, the models suggests that even single morphogen systems can self-organize inside \emph{bi-percolating tissues}, specially when cross and non-linear diffusion are present \cite{vanag_cross-diffusion_2009}.

\section{Turing patterns and instability}
\label{sec:discussion}

\subsection{Spatial heterogeneity and pattern morphologies}
\label{sec:patterns_morphlogies}

\begin{figure}[htbp]
	\centering
	\begin{subfigure}[b]{0.26\textwidth}
		\centering
		\includegraphics[width=\textwidth]{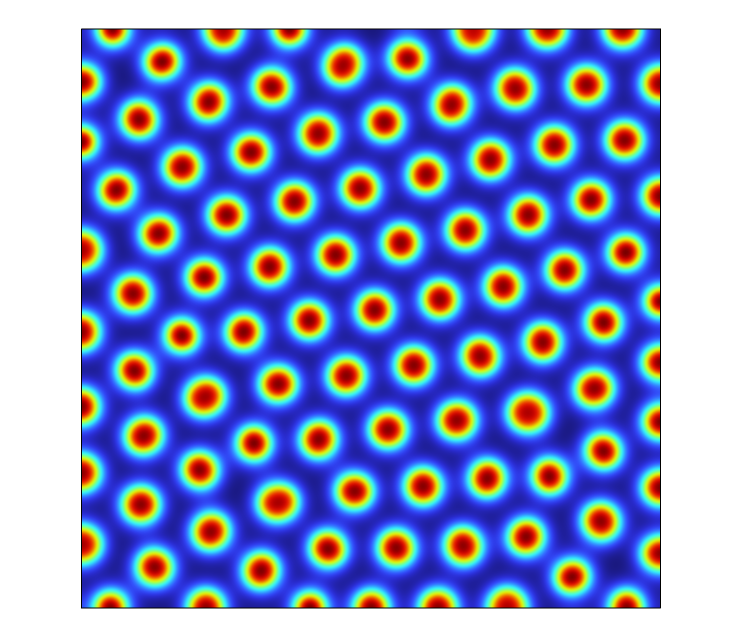}
		\caption{}
		\label{fig:3a}
	\end{subfigure}
	\hfill
	\begin{subfigure}[b]{0.26\textwidth}
		\centering
		\includegraphics[width=\textwidth]{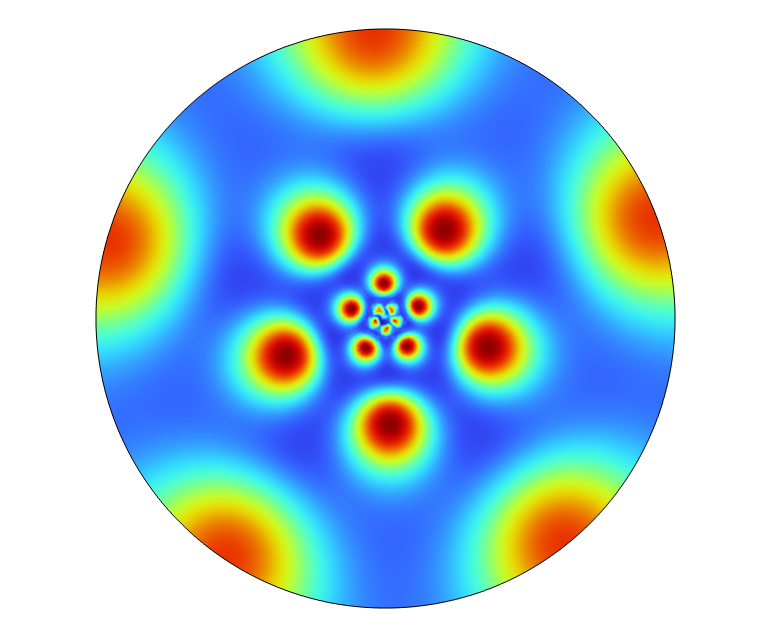}
		\caption{}
		\label{fig:3b}
	\end{subfigure}
	
	\vskip\baselineskip
	
	\begin{subfigure}[b]{0.26\textwidth}
		\centering
		\includegraphics[width=\textwidth]{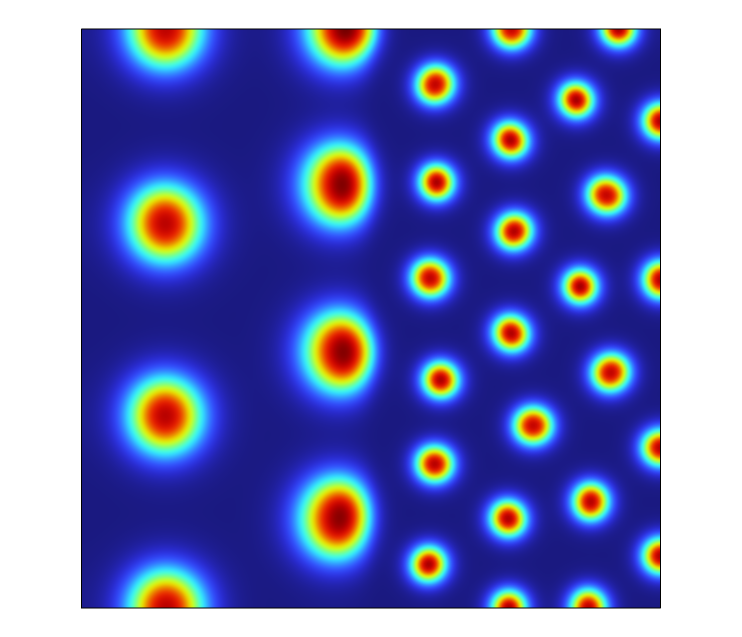}
		\caption{}
		\label{fig:3c}
	\end{subfigure}
	\hfill
	\begin{subfigure}[b]{0.26\textwidth}
		\centering
		\includegraphics[width=\textwidth]{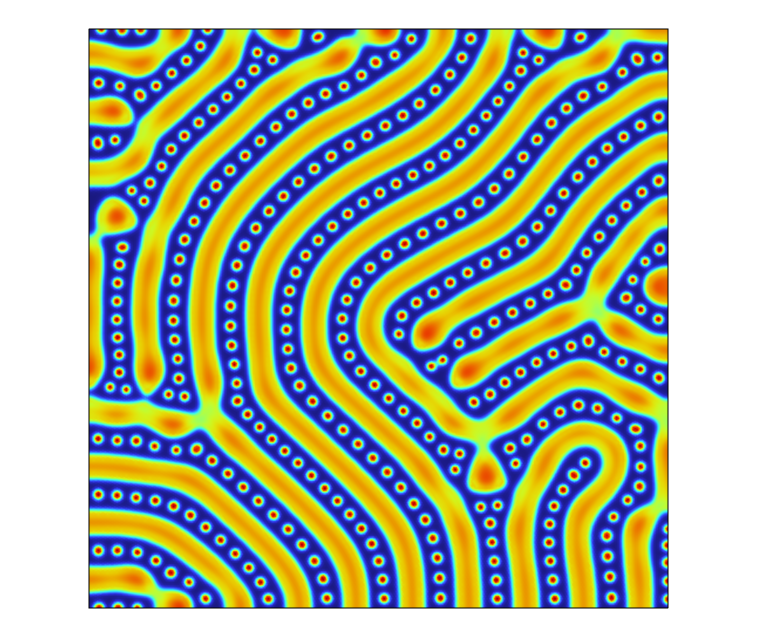}
		\caption{}
		\label{fig:3d}
	\end{subfigure}
	\caption{Examples of Turing patterns with zero flux boundary conditions and different types of spatial heterogeneity. (a) Schnakenberg model on a homogeneous domain as studied in \cite{dufiet_numerical_1992}. (b) BVAM model exhibiting self-similar patterns due to radius-dependent diffusion coefficients \cite{hernandez_self-similar_2017,valdes_lopez_boundary_2024}. (c) Schnakenberg  with anisotropic space-dependent reaction kinetics \cite{van_gorder_pattern_2021}; and (d) Two Brusselator models featuring cubic nonlinear diffusive coupling, as analyzed in \cite{kytta_complex_2007}.  The parameters value for each model and figure are provided in the cited works.}
	\label{fig:2}
\end{figure}

Linearly and non-linearly coupled Turing systems which are particular cases of Eq. (\ref{perc}), have already been used to model pigment patterns in fish skins \cite{barrio_modeling_2009} and superlattice Turing patterns; regular stationary concentration patterns with several characteristic length scales \cite{yang_spatial_2002,kytta_complex_2007}. In these examples, the models represent two thin layers of interacting morphogens that meet at an interface, and are diffusively coupled. In our case on the other hand, the models structure is a consequence of including the cells and their dynamic behavior in the morphogens evolution equations. The fact that these models allow Turing patterns with more than one characteristic length-scale per morphogen pair, is consistent with several results  from research in diffusion-driven instabilities on \emph{heterogeneous domains}. Here,  spatial patterns with several length-scales are the rule, and it has even been shown the possibility of scale-free Turing patterns \cite{hernandez_self-similar_2017}. This opens the possibility for more intricate patterns than usual when more than two morphogens are involved in the process, see Figure (\ref{fig:2}) and references \cite{van_gorder_pattern_2021,calderon-barreto_turing_2022,kytta_complex_2007} for specific examples and a more detailed discussion about this topic.

An interesting feature of the models represented at Equation (\ref{perc}) is that they still allow to include different types of transport (non-linear, cross and anomalous) and spatial heterogeneities (space dependent reaction kinetics/diffusion coefficients) at each physiological domain. Moreover, they allow to explore the effects of physiologically important sources of heterogeneity in Turing 
instabilities and Turing patterns, which are usually absent in these studies, {\it i.e.} space dependent cell properties and cell densities: $\alpha_J = \eta_{\alpha}(x)$, $\beta_{J,k} = \eta_{\tau_J,k}(x)$, $\Delta l = \Delta (x)$, $D_{m,J} = d_{m,J}(x)$, $\rho_c = \rho(x)$ : $x \in \Omega$. Notice also that through the same line of reasoning followed to this point, it is possible to extend the models so to include, for example,  {\em intra-cellular} {\em non-percolating} organelles exchanging molecules with the cytoplasm through their membranes, or  to include advection in the {\em extracellular domain} driven by Starling forces \cite{barrett_ganongs_2019, woodcock_revised_2012}.

\subsection{Turing instability in  physiological contexts}
\label{sec:physiological_contexts}

In this section, we use a {\it bi-percolating model} with normal diffusion at the cell external membranes, to shows that it is possible to have Turing patterns without the usual constrains on the morphogen diffusion coefficients. Additionally, we show that tissue cells  can induce a Turing instability. For this, we use the Schnakenberg reaction mechanism \cite{schnakenberg_simple_1979}, an hypothetical  realistic and simple reaction scheme, widely used to study Turing patterns and their applications:

\begin{equation}\label{sh}
	\begin{split}
		A \overset{k_1}{\longrightarrow} U \\
		U \overset{k_2}{\longrightarrow} A \\
		B \overset{k_3}{\longrightarrow} V \\
		2U+V \overset{k_4}{\longrightarrow} 3U
	\end{split}
\end{equation}

During this chemical process, the concentrations of $A$ and $B$ are maintained constant through what is usually known as the pool chemical assumption \cite{mendez_reactiontransport_2010}. This ensures that the reaction takes place far from thermodynamic equilibrium. In our case, the cells are the ones responsible for this task, and the above mechanism takes the following form:\\
for the intracellular space
\begin{equation}\label{sh1}
	\begin{split}
		Cell \overset{k_1}{\longrightarrow} U, \\
		U \overset{k_2}{\longrightarrow} D,\\
		Cell \overset{k_3}{\longrightarrow} V, \\
		2U+V \overset{k_4}{\longrightarrow} 3U;
	\end{split}
\end{equation}
for the extracellular space
\begin{equation}\label{sh2}
	\begin{split}
		U \overset{k_2^*}{\longrightarrow} D,\\
		2U+V \overset{k_4^*}{\longrightarrow} 3U,
	\end{split}
\end{equation}
where $D$ denotes a reaction product that does not participate in the process. \\
Now, if the reaction occurs in a {\it bi-percolating tissue}, the morphogens evolution equations are: 

\begin{equation}
	\label{shm}
	\begin{aligned}
		\frac{\partial \rho_{U,i}}{\partial t}
		={}&d_{U,i}\nabla^2\rho_{U,i}
		+k_1-k_2\rho_{U,i}
		+\frac{\rho_c s_c d_{U,m}}{\Delta l}
		(\rho_{U,o}-\rho_{U,i})
		\\
		&+\eta_{U,i}(\rho_{U,i},\rho_{V,i}),
		\\[2mm]
		\frac{\partial \rho_{U,o}}{\partial t}
		={}&d_{U,o}\nabla^2\rho_{U,o}
		-k_2^*\rho_{U,o}
		+\frac{\rho_c s_c d_{U,m}}{\Delta l}
		(\rho_{U,i}-\rho_{U,o})
		\\
		&+\eta_{U,o}(\rho_{U,o},\rho_{V,o}),
		\\[2mm]
		\frac{\partial \rho_{V,i}}{\partial t}
		={}&d_{V,i}\nabla^2\rho_{V,i}
		+k_3
		+\frac{\rho_c s_c d_{V,m}}{\Delta l}
		(\rho_{V,o}-\rho_{V,i})
		\\
		&+\eta_{V,i}(\rho_{U,i},\rho_{V,i}),
		\\[2mm]
		\frac{\partial \rho_{V,o}}{\partial t}
		={}&d_{V,o}\nabla^2\rho_{V,o}
		+\frac{\rho_c s_c d_{V,m}}{\Delta l}
		(\rho_{V,i}-\rho_{V,o})
		\\
		&+\eta_{V,o}(\rho_{U,o},\rho_{V,o}),
		\\[2mm]
		\eta_{U,i}
		={}&k_4\rho_{U,i}^{2}\rho_{V,i},
		\qquad
		\eta_{U,o}
		=
		k_4^*\rho_{U,o}^{2}\rho_{V,o},
		\\
		\eta_{V,i}
		={}&-\eta_{U,i},
		\qquad
		\eta_{V,o}
		=
		-\eta_{U,o}.
	\end{aligned}
\end{equation}

For a Turing instability to exits, it is necessary that the homogeneous steady state of interest is linearly stable in the absence of diffusion, otherwise it must be linearly unstable against perturbations with finite characteristic length-scales. For this model, there are at least two homogeneous steady states of interest: $(\rho_{U,i},\rho_{U,o},\rho_{V,i},\rho_{V,o})=(k_1+k_2,0,k_2/(k_1+k_2)^2,0)$ and  $(\rho_{U,i},\rho_{U,o},\rho_{V,i},\rho_{V,o})=(\rho_{U}^*,\rho_{U}^*,\rho_{V}^*,\rho_{V}^*)$. The first correspond to the solution of Eq.(\ref{shm}) when $d_{U,i}=d_{U,o}=d_{V,i}=d_{V,o}=d_{U,m}=d_{V,m}=0$. The second solution, is obtained for $d_{U,i}=d_{U,o}=d_{V,i}=d_{V,o}=0$ and $d_{U,m},d_{V,m}>0$. For reasons of space, in what follow we will focus on the second case, which is also the more standard version of a Turing analysis.

Making the system in Eq.(\ref{shm}) dimensionless via Murray's proposed change of variables \cite{murray_mathematical_2003}, and assuming equal reaction rates in both intracellular and extracellular media for simplicity (though this assumption is not strictly necessary), yields:

\begin{equation}
	\begin{aligned}
		\frac{\partial u_i}{\partial t} &=  d_{u,i} \nabla^2 u_i + \gamma\left(a - u_i + u_i^2 v_i + c(u_o - u_i)\right) \\
		\frac{\partial v_i}{\partial t} &=  d_{v,i} \nabla^2 v_i + \gamma\left(b - u_i^2 v_i + e(v_o - v_i)\right) \\
		\frac{\partial u_o}{\partial t} &=  \nabla^2 u_o + \gamma\left(u_o^2 v_o - u_o + c(u_i - u_o)\right) \\
		\frac{\partial v_o}{\partial t} &=  d \nabla^2 v_o + \gamma\left(-u_o^2 v_o + e(v_i - v_o)\right)
	\end{aligned}
	\label{ec:dimensionless_system}
\end{equation}

\begin{figure}[htbp]
	\centering
	\begin{subfigure}[b]{0.48\textwidth}
		\centering
		\includegraphics[width=\textwidth]{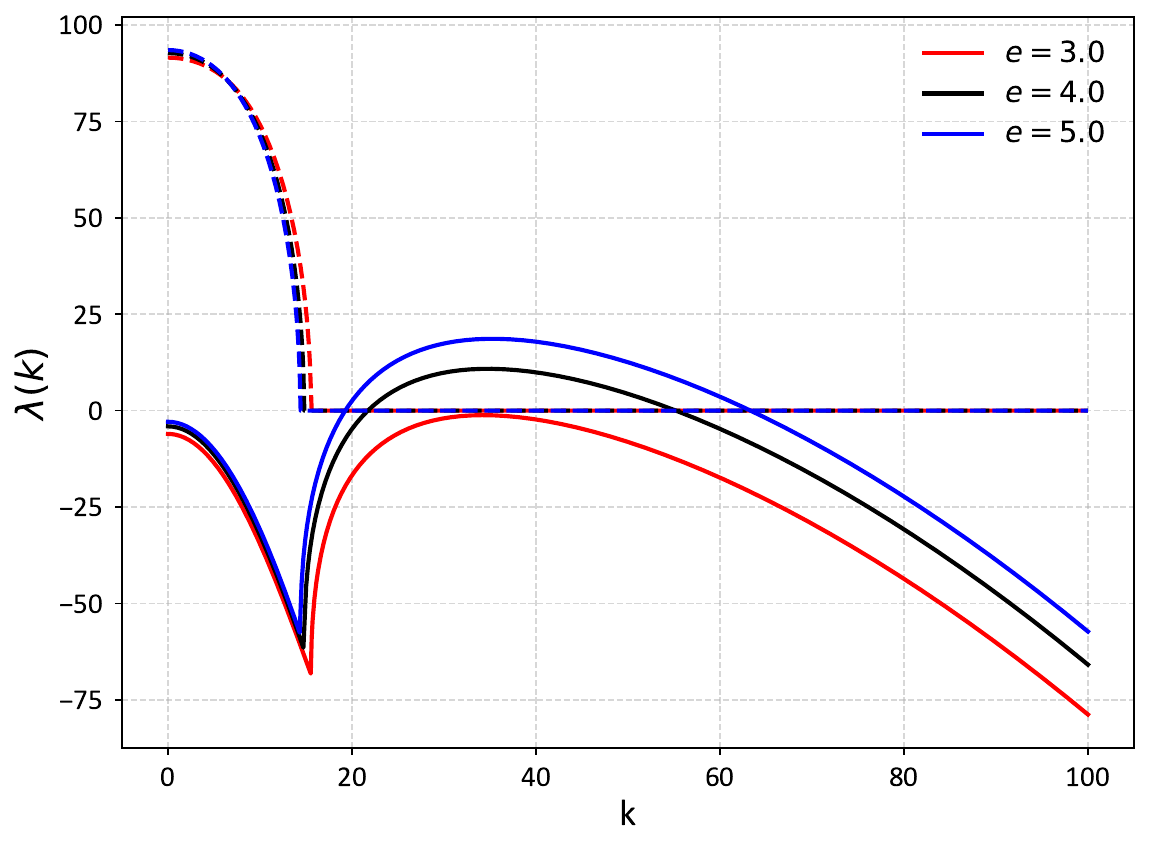}
		\caption{$d=1$}
		\label{fig:4a}
	\end{subfigure}
	\hfill
	\begin{subfigure}[b]{0.48\textwidth}
		\centering
		\includegraphics[width=\textwidth]{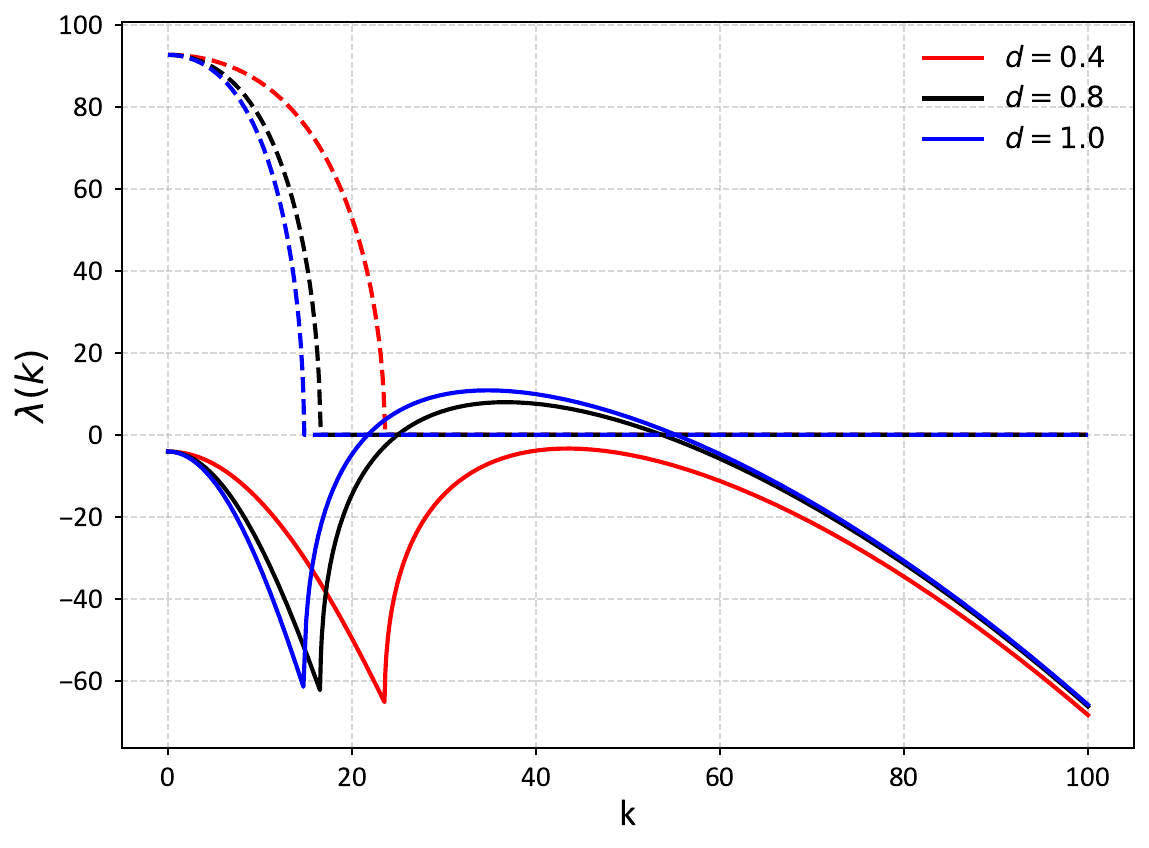}
		\caption{$e=4$}
		\label{fig:4b}
	\end{subfigure}
	
	\caption{Real (continuous lines) and imaginary (dashed lines) parts of the dispersion relation for different values of relevant model parameters. For both subfigures, the parameters were $a=0.07$, $b=1.34$, $c=0.05$, $\gamma=100$, $d_{u,i}=0.01$, $d_{v,i}=0.01$, and the fixed point was calculated numerically, coinciding to four significant figures to $\left( u_i^*,v_i^*,u_o^*,v_o^* \right) = \left( 1.342, 0.741, 0.067, 0.740 \right)$.}
	\label{fig:4}
\end{figure}

\begin{figure}[htbp]
	\centering
	\includegraphics[width=0.48\textwidth]{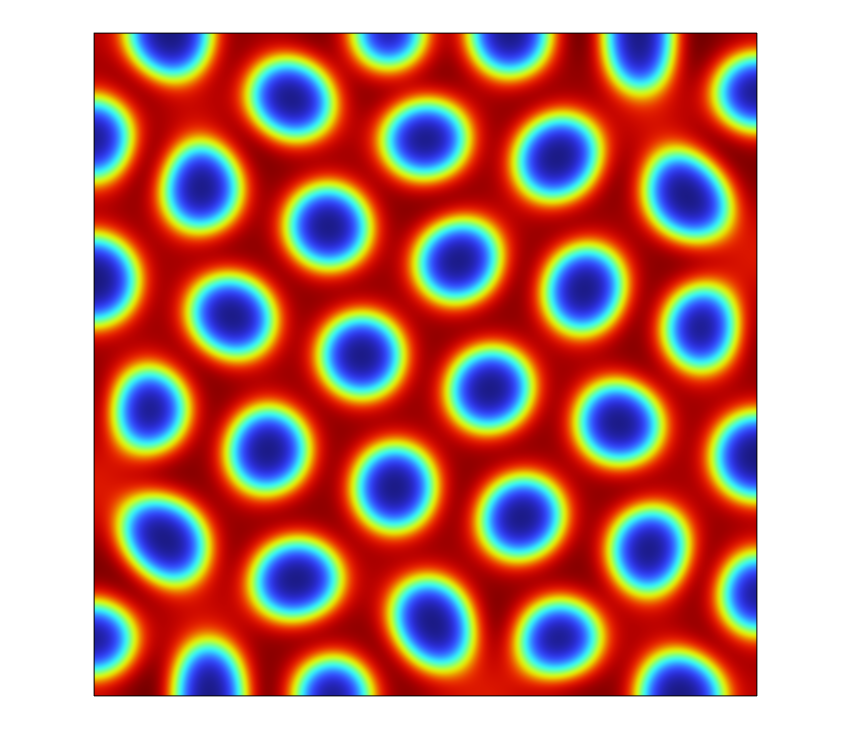}
	\caption{Example of a Turing pattern under conditions where the activator diffuses faster than the inhibitor ($d<1$). The parameters used in the simulation were $a=0.07$, $b=1.34$, $c=0.05$, $e=4$, $d=0.8$, $\gamma=100$, $d_{u,i}=0.01$, $d_{v,i}=0.01$. Zero-flux boundary conditions were imposed, and the fixed point is $\left( u_i^*,v_i^*,u_o^*,v_o^* \right) = \left( 1.34288, 0.74123, 0.06712, 0.74039 \right)$.} 
	\label{fig5}
\end{figure}

Figure (\ref{fig:4}) presents dispersion relations as functions of relevant model parameters. Note that both the membrane permeability to the inhibitor ($e$) and the ratio of diffusion coefficients in the extracellular medium ($d$) act as bifurcation parameters. These results are significant for two reasons. Regarding membrane permeability (Figure (\ref{fig:4a})), this parameter is controllable by the cellular machinery, providing a mechanism to regulate Turing pattern formation beyond morphogen production inside cells. Moreover, concerning the diffusion coefficient ratio (Figure (\ref{fig:4b})), Turing patterns can emerge even when the activator diffuse faster than the inhibitor (see Figure (\ref{fig5})), which is impossible in classical homogeneous models. We found this phenomenon of Turing patterns formation when $d<1$ is also possible if chemical reactions between morphogens are suppressed in the extracellular medium, i. e $\eta_{U,o} = \eta_{V,o} = 0$, a reasonable assumption given that extracellular environments are known to differ substantially from intracellular conditions.

The previous results suggest that Turing pattern formation may be more prevalent in biological systems than currently believed. By expanding the parameter space where patterns are possible, our tissue-structured model reduces the stringent conditions typically required, underscoring the importance of incorporating spatial heterogeneity to capture the complexity of Turing-type patterns in biological contexts. Moreover, although not shown here, the possibility of generating spot and stripe$-$like structures enables our model to be applied in explaining phenomena such as digit and feather patterning, among others reported in the literature \cite{raspopovic_digit_2014,ho_feather_2019}, but with the advantage of being grounded in first principles.  

As a final observation, we would like to highlight the potential role of molecule $D$ in the mechanism proposed in Eqs. (\ref{sh1}) and (\ref{sh2}), which has been neglected thus far. Although this substance does not participate directly in the activator-inhibitor morphogen reaction, it may modify the properties of intracellular and/or extracellular media (e.g., through changes in viscosity, molecular crowding, etc.), thereby facilitating the formation of Turing patterns, similar to what was reported in \cite{aizawa_theory_2024}. Another biologically plausible scenario is that $D$ modifies morphogen transport across the cell membrane, which, as we emphasized previously, also functions as a bifurcation parameter. In both scenarios, the parameters of the reaction-diffusion system would no longer be constant but would depend dynamically on the system's state. We therefore believe that under these conditions, the system could self-tune toward the threshold of Turing instability without requiring external fine-tuning, which is consistent with the phenomenon of self-organized criticality \cite{bak_self-organized_1987}. 

While this remains a hypothesis, such a mechanism has been previously highlighted as a phenomenon of interest \cite{yang_turing_2003,mercker_mechanochemical_2015} and, together with the expansion of the parameter space in which pattern formation is possible, it is addressed one of the Turing's theory's most common criticisms: the \emph{fine-tuning problem}. Furthermore, since the model's parameters would become space- and time-dependent fields, non-stationary patterns, domains with variable wavelengths, or quasi-self-similar structures could emerge, something that is not common in the classical Turing model with constant parameters.

\section{Conclusions}
\label{sec:conclusions}

The physiologically grounded morphogenetic models developed in this 
work, reveal a great capacity of living tissues to generate rich and complex dynamic behavior that is under cellular control, and occur under weaker physico-chemical non-equilibrium constrains than the ones required by classical theory \cite{murray_mathematical_2003,keener_mathematical_2009,turing_chemical_1952}. Such capacity is a direct consequence of including the cell internal dynamics and their associated spatial heterogeneities and selective permeabilities in the models mathematical structure.

Being derived from conservation laws, cellular transport mechanisms, and reaction kinetics, the models parameters have straightforward physiological interpretation, opening the door to experimental predictions and validations. More importantly, the models and ideas herein discussed, offer a new perspective for studying emergence of Turing-like spatial patterns beyond the classical homogeneous framework, highlighting how realistic physiological scenarios can weaken classical constraints and lead to a broader spectrum 
of morphogenetic phenomena. 

The models exploration and extension to more realistic scenarios, could contribute to ongoing efforts in the fields of developmental biology and synthetic pattern and tissue engineering, providing a minimal, yet versatile platform to understand, and potentially control pattern formation processes in and during tissue development. In other contexts and as more distant applications, the models could be useful for describing emergent collective behaviors in biofilms and excitable tissues. Also, notice that the amplification of morphogens dynamic possibilities inside tissues or cell aggregates, implies an expansion of their possible collective responses to environmental challenges, and an enlargement of its adjacent possible \cite{kauffman_third_2023}, providing more favorable conditions for the aggregate to find successful strategies/mechanisms for multicellular transitions in different species during biological evolutionary history \cite{ruiz-trillo_origins_2007,grosberg_evolution_2007}.

\begin{acknowledgments}
Alejandro Vald{\'e}s L{\'o}pez gratefully acknowledges the financial support provided by the Secretar{\'i}a de Ciencia, Humanidades, Tecnolog{\'i}a e Innovaci{\'o}n (SECIHTI) under Grant No. 4016891.
\end{acknowledgments}



%

\end{document}